\documentstyle[12pt]{article}
\begin{document}
\title{Space Time Quantization and the Big Bang$^*$}
\author{B.G. Sidharth\\
Centre for Applicable Mathematics \& Computer Sciences\\
B.M. Birla Science Centre, Adarsh Nagar, Hyderabad - 500 063 (India)}
\date{}
\footnotetext{Email:birlasc@hd1.vsnl.net.in}
\maketitle
\begin{abstract}
A recent cosmological model is recapitulated which deduces the correct
mass, radius and age of the universe as also the Hubble constant and
other well known apparently coincidental relations. It also predicts
an ever expanding accelerating universe as is confirmed by latest
supernovae observations. Finally the Big Bang model is recovered
as a suitable limiting case.
\end{abstract}
\section{Introduction}
Our starting point is a recent cosmological model\cite{r1,r2,r3}, in which the
well known micro physical constants\cite{r4}, namely the mass or Compton
wavelength and charge of a typical elementary particle as also the Planck
constant and the velocity of light, are used along with the particle number
$N \sim 10^{80}$ as the sole large scale or cosmological parameter. Then
one can deduce from the theory the correct values of the mass, the radius
and age of the universe, as also the Hubble constant and the cosmological
constant. The relation between the mass of the pion and the Hubble constant,
described by Weinberg as mysterious\cite{r5} as also the large number
"coincidences" which lead to the Dirac large number hypothesis also follow
as a consequence. In addition the model predicts that the universe will
accelerate and expand for ever, as has been confirmed by several recent
independent observers\cite{r6,r7,r8}.\\
The object of this paper is to show how the standard Big Bang theory can
be recovered from the above model.
\section{The New Cosmological Model and the Big Bang}
While readers are referred to references\cite{r1,r2,r3} for details, it may be
mentioned that there are two main concepts underlying the model. Firstly,
particles, typically pions are created from the background Zero Point Field
by fluctuations within space time intervals $(l, \tau)$, the corresponding
Compton wavelength and Compton time. The idea here is that the usual space
time points that are used are only classical approximations as pointed out
by Wheeler\cite{r9}. In a strict Quantum Mechanical context, this is no
longer valid, and one has to consider, as argued in detail, the Compton
wavelength and Compton time as the minimum physical space time intervals.
This "quantization" is reminiscent of the chronon\cite{r10,r11}.
In this connection it may be pointed out, for example, that in the Dirac
theory of the electron, a point electron has the velocity of light-- it is
only on averaging over the above space time intervals do we recover physical
subluminal particles or equivalently, the position operator becomes Hermitian\cite{r12}.
Infact it was shown that there is a holistic basis for this, what was called
a micro-macro nexus.\\
Secondly if at a given epoch there are $N$ particles in the universe,
$\sqrt{N}$ particles are fluctuationally created. Infact as shown in
detail elsewhere by Hyakawa (cf. for example\cite{r13}), this leads to
perfectly meaningful results, which are otherwise inexplicable.\\
To recapitulate (cf. references\cite{r1,r2,r3} for details), some of the
relations which are deduced and consistent with observation are:
$$M = Nm,$$
$$R = \frac{GM}{c^2}$$
$$\sqrt{N} = \frac{2mc^2}{\hbar} T$$
where $m$ is the pion mass, $M$ is the mass of the universe, $R$ its radius
and $T$ is its age. The last equation follows from the relation
\begin{equation}
\frac{dN}{dt} = \frac{\sqrt{N}}{\tau} = \frac{mc^2}{\hbar}\sqrt{N}\label{e1}
\end{equation}
in which we use the fact that the minimum time interval is the Compton time
$\tau$ of a typical elementary particle, namely the pion.\\
We further deduce that
$$H = \frac{Gm^3c}{\hbar^2},$$
$$\wedge \sim H^2$$
where $H$ is the Hubble constant and $\wedge$ is the cosmological constant.
Both these relations are consistent with observation. The equation for the
Hubble constant can be rewritten as
$$m = (\frac{\hbar^2 H}{Gc})^{1/3}$$
which is a mysterious relation refered to by Weinberg. It is now a consequence
of the theory and expresses the holistic feature alluded to, above.\\
Another hitherto empirical relation that can be deduced in the above model is
\begin{equation}
\frac{e^2}{Gm^2} \approx \sqrt{N}\label{e2}
\end{equation}
We now observe that we should be able to recover standard theory in the
classical approximation in which the minimum space time intervals ($l, \tau)
\to 0$.\\
Indeed, we can see immediately from equation (\ref{e1}) that as $\tau \to 0,
\frac{dN}{dt} \to \infty$ corresponding to a singular creation of the $N$ particles of the
universe.\\
More accurately, as has been pointed out by several authors\cite{r14,r15}, the
minimum space time intervals are at the Planck scale. Indeed as has been noted
\cite{r16,r17}, the Planck scale is the extreme limit where Quantum Mechanics
meets the purely classical General Relativity. One easy way to see this is that
for a Planck mass $m_P, \sim 10^{-5}gms$, we have,
$$\frac{Gm_P}{c^2} = \hbar/m_Pc$$
where the left side represents the purely classical Schwarzchild Black Hole
radius and the right side the Quantum Mechanical Compton wavelength.\\
Another way of looking at this is that for the Planck mass we have instead
of equation (\ref{e2}), the relation
$$\frac{Gm_P^2}{e^2} \approx 1,$$
which shows that all the energy is gravitational.\\
We now use the fact that our minimum space time intervals are $(l_P, \tau_P)$,
the Planck scale, instead of $(l, \tau)$ of the pion, as above.\\
With this new limit, it can be easily verified that the total mass in the volume
$\sim l^3$ is given by
$$\rho_P \times l^3 = M$$
where $\rho_P$ is the Planck density and $M$ as before is the mass of the
universe.\\
Moreover the number of Planck masses in the above volume $\sim l^3$ can
easily be seen to be $N' \sim 10^{60}$. However, it must be remembered that
in the physical time period $\tau$, there are $10^{20}$ (that is $\frac{\tau}
{\tau_P})$ Planck life times. In other words the number of Planck particles
in the physical interval $(l, \tau)$ is $N \sim 10^{80}$.\\
That is from the typical physical interval $(l, \tau)$ we recover the entire
mass and also the entire number of particles in the universe, as in the Big
Bang theory. This also provides the explanation for the above puzzling
relations.\\
That is the Big Bang theory is a characterization of the new model in the
classical limit at Planck scales.

\end{document}